\documentclass[%
 reprint, superscriptaddress,
nofootinbib,
 amsmath,amssymb,
 aps,
]{revtex4-2}
\pdfoutput=1 %

\usepackage{braket}
\usepackage{amsmath,amssymb,amsfonts}
\usepackage{amsthm}
\usepackage{mathtools}
\usepackage{graphicx}
\usepackage{xcolor}
\usepackage{hyperref}
\usepackage{framed}

\usepackage[capitalize,nameinlink]{cleveref}
\usepackage{comment}

\newcommand{\be}{\begin{equation}}\newcommand{\ee}{\end{equation}}\newcommand{\ba}{\begin{eqnarray}}\newcommand{\ea}{\end{eqnarray}}\newcommand{\ban}{\begin{eqnarray*}}\newcommand{\ean}{\end{eqnarray*}}
\newcommand{\avg}[1]{\left\langle #1 \right\rangle}
\newcommand{\one}{\openone}

\newtheorem{lemma}{Lemma}

\theoremstyle{definition}

\theoremstyle{remark}

\def\mE{\mathcal{E}}

\def\openone{\mathds{1}}

\newcommand{\sandwich}[3]{\langle #1 |#2|#3\rangle}
\newcommand{\ketbra}[2]{|#1\rangle\!\!\!\; \langle #2 |}
\newcommand{\ketbrasame}[1]{|#1\rangle\!\!\!\; \langle #1 |}
\newcommand{\tr}[1]{\operatorname{Tr}\!\left[#1\right]}

\newcommand{\HS}{{\mathcal{H}}}

\usepackage{dsfont}
\usepackage{enumitem}

\newcommand{\Tr}{\mathrm{Tr}}

\newcommand{\map}[1]{\mathcal{#1}}

\renewcommand{\geq}{\geqslant}
\renewcommand{\leq}{\leqslant}
\renewcommand{\ge}{\geqslant}

\begin{document}

\title{A quantum entropy production operator}%

\author{Ge Bai}
\email[Corresponding author: ]{gebai@hkust-gz.edu.cn}
\affiliation{Thrust of Artificial Intelligence, Information Hub, The Hong Kong University of Science and Technology (Guangzhou), Guangzhou 511453, China}
\affiliation{Centre for Quantum Technologies, National University of Singapore, 3 Science Drive 2, Singapore 117543}

\author{Francesco Buscemi}
\affiliation{Department of Mathematical Informatics, Nagoya University, Furo-cho Chikusa-ku, Nagoya 464-8601, Japan}

\author{Valerio Scarani}
\affiliation{Centre for Quantum Technologies, National University of Singapore, 3 Science Drive 2, Singapore 117543}
\affiliation{Department of Physics, National University of Singapore, 2 Science Drive 3, Singapore 117542}

\date{\today}%

\begin{abstract}

We introduce a fully quantum notion of entropy production based on the non-commutative extension of the classical log-ratio between forward and reverse processes. Given a pair of quantum objects associated with the forward and reverse descriptions, we define a Hermitian entropy-production operator whose expectation value is non-negative and equal to the Belavkin--Staszewski relative entropy. The operator satisfies exact integral and detailed fluctuation theorems without requiring commutativity.

We then specialize this construction to the case in which the forward process is described by a single quantum channel and the reverse process is defined inferentially, through Bayesian retrodiction relative to a prior state, with the Petz transpose map as the Bayesian inverse. In this setting, the relevant quantities can be evaluated explicitly, leading to a number of natural structural and physical properties. The framework recovers the classical formula in the commutative limit, yields explicit expressions for the average entropy production, and clarifies where the fully quantum case departs from standard thermodynamic expectations.

\end{abstract}

\maketitle

\section{Introduction}

In classical stochastic thermodynamics, once the relevant states $m$ of the system under study have been identified, one can assign values of the relevant thermodynamical quantities to individual realizations of the process. Macroscopic quantities are then obtained by averaging over trajectories. In many situations, the microscopic value of an entropy contribution, such as the total entropy production or its irreversible part, takes the log-ratio form
\begin{align}\label{eq1}
    s(\textbf{m})=\log\,\frac{P_F(\textbf{m})}{P_R(\textbf{m})}\,.
\end{align}
Here, $P_F$ is the probability distribution of the trajectories of the forward process, while $P_R$ refers to a suitable notion of reverse process. The importance of this expression is well known. Van den Broeck called \cref{eq1} ``what we believe to be the single most important expression in stochastic thermodynamics'' \cite{van-den-broeck-stat-thermo-introduction}. At the same time, the meaning of the reverse process is not unique. Depending on the physical setting, it may be defined by evaluating $P_F$ on the inverse trajectory \cite{Seifert-2005}, by considering the reverse trajectory under an inverted external drive \cite{crooks-theorem}, or by yet other constructions \cite{Esposito2010}. Similar expressions may also hold for ensembles of trajectories, for instance after grouping together all those that share the same initial and final points. Various reviews have been devoted to these matters \cite{esposito-RMP-review,jarzynski-review-2011,Seifert_2012,review_vdbEsp15,Marsland_2018}.

This flexibility in the choice of reverse process is both a strength and a source of ambiguity. On the one hand, log-ratio expressions have a fundamental flavor: entropy, as a measure of irreversibility, appears as a statistical comparison between the physical process and a suitable reverse. On the other hand, different notions of reverse process are tied to different protocols and different physical questions. Many of the usual constructions are operationally motivated, in the sense that the reverse process is specified by how one intends to reverse the dynamics or the driving. In this work, we will eventually specialize to an inferential notion of reverse process, constructed from Bayesian retrodiction. This viewpoint does not start from a reversed experimental protocol, but from the task of inferring past states from present information.

The formal consequences of log-ratio expressions are also very useful. The most straightforward consequence is that the average entropy is non-negative. Indeed, we have \cite{kawai}
\begin{align}
    S\,\coloneqq\,\avg{s}_F=D(P_F||P_R)\,\geq 0 \,,
\end{align}
which is the Kullback-Leibler divergence---a measure of the distinguishability of $P_F$ and $P_R$ known to be non-negative. But knowledge of the individual $s$ opens the door to deriving fluctuation theorems. In particular, it is easy to derive the detailed fluctuation theorem (see e.g.~\cite{AwBS})
\begin{align}\label{detailed_c}
    P_R(-s)=e^{-s}\,P_F(s)\,,
\end{align}
whence the normalisation of $P_R$ leads to the integral fluctuation theorem
\begin{align}\label{integral_c}
    \avg{e^{-s}}_F=1\,.
\end{align}
From the latter, $\avg{s}\geq 0$ follows by Jensen's inequality.

In quantum thermodynamics, entropy production remains one of the key quantities. As a canonical example, when the coupling with the thermal bath is weak, the irreversible entropy production of a Gibbs channel $\mathcal{E}$ is given by
\begin{align}\label{eq2}
    \Sigma=\,D(\rho||\rho^{\textrm{eq}})\,-\,D(\mathcal{E}[\rho]||\rho^{\textrm{eq}})
\end{align}
where $D$ is the Umegaki relative entropy defined as
\begin{align}
    \label{umegaki} D(\rho\|\sigma):=\Tr[\rho(\log\rho-\log\sigma)],\quad\sigma>0\;,
\end{align}
and $\rho^{\textrm{eq}}$ is the Gibbs state of the system \cite{EspositoNJP,DeffnerLutz11}. This is the result of a calculation analogous to the classical rewriting of an entropy difference in terms of joint input-output statistics, but the passage to a log-ratio form is blocked by two obstacles: one would need a quantum analog of such joint process statistics, which is not an obvious object because of no-cloning; and the identity $\log\rho-\log\rho'=\log(\rho/\rho')$ holds only iff $[\rho,\rho']=0$ (if this does not hold, the r.h.s.~itself is ill-defined).

When it comes to quantum fluctuation theorems \cite{campisi-haenggi-review-2011,landi-RMP-review}, most examples use the so-called two-point measurement schemes, variations of the pioneering insight of Tasaki \cite{tasaki2000jarzynski}. For these examples, it is proved that the fluctuating quantity (the difference of the two measured energies) cannot be interpreted as the value of a ``work observable'' \cite{heisenberg-work-op,Hovhannisyan2024energyconservation,rubino2024revisingquantumworkfluctuation}. Other fluctuation theorems have been proposed for the observed outcomes of monitored quantum thermal processes \cite{manzano-PRX,micadei2020quantum} and for general quantum channels \cite{kwon-kim,BS21}, but without identifying an underlying quantum observable that is fluctuating. Operator expressions that generalize fluctuation theorems have been proposed for thermal processes \cite{Alhambra16,aberg-quantum-fluct}. A recent formal proposal to replace probabilities with quasi-probabilities in \cref{eq1} results in individual entropies, and even average ones, that assume complex values \cite{twesh}.

Our aim is to investigate, in the quantum setting, how the classical log-ratio structure of entropy production can be extended beyond commutativity. In Section \ref{seclog}, we first address this question at a general formal level, by introducing an operator $\sigma$ that satisfies the main formal properties of the classical counterpart $s$: it is Hermitian, its average $\Sigma=\avg{\sigma}$ is non-negative, and it satisfies a detailed fluctuation theorem together with the corresponding integral one. In Section \ref{secsot}, we then give physical content to this operator by specializing to processes described by a single CPTP map and to an inferential notion of reverse process constructed from Bayesian retrodiction. The technical challenge there is to identify quantum objects that play the role of forward and reverse input-output statistics and yield reasonable expressions for the average entropy $\Sigma$.

\section{A fully quantum analog of the thermodynamic log-ratio}
\label{seclog}

Before turning to a concrete construction of the forward and reverse processes, we isolate the formal problem that lies at the core of the classical approach. In the classical case, entropy production is defined as a log-ratio of the probability distributions of the forward and reverse trajectories. Here we ask what should count as a fully quantum version of that structure, once one is given quantum objects playing the role of such trajectory distributions.

We therefore assume that there exist quantum analogs $Q_F$ and $Q_R$ of the classical probability distributions $P_F(\mathbf{m})$ and $P_R(\mathbf{m})$ over the trajectories. We take these objects to be Hermitian, positive semi-definite, and unit trace, that is, the analog of quantum states. Our goal in this section is to identify a non-commutative counterpart of the log-ratio \eqref{eq1} that preserves the main formal consequences of the classical expression, namely the non-negativity of the average entropy production and the validity of integral and detailed fluctuation theorems. We define the fully quantum analog of the log-ratio \eqref{eq1} as the operator
\begin{align}\label{eq:def-operator}
 \sigma[Q_F,Q_R]\coloneqq
 \log \left\{\sqrt{Q_F}\,Q_R^{-1}\,\sqrt{Q_F}\right\}\,.
\end{align}
We implicitly assume that $Q_R$ is invertible on the support of $Q_F$, so as not to run into division by zero, as always happens when considering entropic quantities. The logarithm must likewise be restricted to the support of its argument.

The specific form of \cref{eq:def-operator} is not the only definition that reduces to the classical entropy production formula \ref{eq1} when operators $Q_F$ and $Q_R$ commute. Alternatives like $\log \left\{Q_R^{-1/2}Q_FQ_R^{-1/2}\right\}$ and $[\log Q_F-\log Q_R]$ also satisfy this basic criterion. However, these alternative definitions fail to satisfy the expected fluctuation theorems. In particular, neither of them always satisfies \cref{qjar} below. This makes \cref{eq:def-operator} a distinguished non-commutative extension of the classical log-ratio.

We now turn to the formal properties of this operator. First of all, its average with respect to $Q_F$ is non-negative by construction:
\begin{align}
\avg{\sigma[Q_F,Q_R]}_F&\coloneqq \Tr\Big\{Q_F\;\sigma[Q_F,Q_R]\Big\}\nonumber\\
&=\textrm{Tr}\left[ Q_F \log \sqrt{Q_F}\,Q_R^{-1}\,\sqrt{Q_F}\right]\nonumber\\
    &\eqqcolon D_{\textrm{BS}}(Q_F\|Q_R)\,\geq 0\label{avq1}
\end{align}
where $D_{\textrm{BS}}$ is the \textit{Belavkin--Staszewski relative entropy} \cite{belavkin1982c,matsumoto2015new}. This relative entropy is an upper bound on Umegaki's relative entropy and, for commuting operators, coincides with the KL-divergence. We refer the reader to Refs.~\cite{tomamichel2015quantum,khatri2020principles} for further properties and relations to other divergences. Moreover, $D_{\textrm{BS}}(Q_F\|Q_R)=0$ if and only if $Q_F=Q_R$.

A second key property is that \cref{eq:def-operator} satisfies a fully quantum analog of the integral fluctuation theorem:
\begin{align}\label{qjar}
    \avg{e^{-\sigma[Q_F,Q_R]}}_F&=\textrm{Tr}\left[Q_F\; e^{-\sigma[Q_F,Q_R]}\right]\nonumber\\
    &=\Tr[Q_R]=1\,,
\end{align}
whenever $Q_F$ and $Q_R$ are supported in the same subspace. Just as in the classical case \eqref{integral_c}, this relation can be read formally as the normalization of the reverse process.

We next consider the detailed fluctuation theorem. Since the operator \eqref{eq:def-operator} is Hermitian, its eigenvalues provide the candidate values of entropy production:
\begin{align}
    \sigma[Q_F,Q_R]\ket{f_k}&=\,s_k\ket{f_k}\,. \label{eq:fk}
\end{align}
Assuming all the eigenvalues are discrete and non-degenerate, the corresponding statistics in the forward process are given by
\begin{align}
    \mathbb{P}_F(s_k)&=\,\sandwich{f_k}{Q_F}{f_k}\,. \label{eq:epfwd}
\end{align}
The corresponding operator for the reverse process is obtained by exchanging the roles of $Q_R$ and $Q_F$. In contrast to the classical case, the operator identity $\sigma[Q_R,Q_F]=-\sigma[Q_F,Q_R]$ does not hold in general. More precisely, the relation holds for the eigenvalues, which are the same with opposite signs, but the eigenvectors are in general different (see \cref{app:crooks}):
\begin{align}
    \sigma[Q_R,Q_F]\ket{r_k}&=\,-s_k\ket{r_k}\,. \label{eq:rk}
\end{align}
The corresponding statistics on the reverse process are thus given by
\begin{align}
    \mathbb{P}_R(-s_k)&=\,\sandwich{r_k}{Q_R}{r_k} \,. \label{eq:eprev}
\end{align}

Although this is not immediate, by carefully studying the singular value decomposition of $Q_F^{1/2}\,Q_R^{-1/2}$, one can prove that the probability distributions defined in \cref{eq:epfwd,eq:eprev} obey the relation
\begin{align} \label{crooksc}
    \mathbb{P}_R(-s_k)&=\,e^{-s_k}\,\mathbb{P}_F(s_k)
\end{align}
which is the same as the classical detailed fluctuation theorem \eqref{detailed_c}. The proof can be found in \cref{app:crooks}. Thus, not only a detailed fluctuation theorem can be written down, but the very same exponential relation survives in a setting where forward and reverse operators do not commute.

Taken together, these observations show that \cref{eq:def-operator} preserves an unexpectedly large portion of the classical log-ratio structure. It is Hermitian, its average is a non-negative relative entropy, and it satisfies exact integral and detailed fluctuation theorems. None of these properties is automatic in the non-commutative setting, and retaining all of them at once is highly restrictive. In this sense, the operator $\sigma[Q_F,Q_R]$ seems to be a particularly natural and robust fully quantum analog of its classical counterpart.

\section{Inferential quantum entropy production for quantum channels}
\label{secsot}

All the properties of $\sigma[Q_F,Q_R]$ derived in the previous section rely only on the assumption that the operators $Q$ are Hermitian, positive semidefinite, and of unit trace. However, we have not yet provided an explicit construction. In this section we present such a construction, based on Bayesian inversion. To give these formal results physical content, we must identify operators of this kind that correspond to concrete physical processes and that reproduce familiar thermodynamic expressions, at least in appropriate limiting regimes. We do so here for a simple yet relevant class of processes: those described by a single completely positive trace-preserving (CPTP) map $\mathcal{E}$.

For the forward process, the relevant data are the initial state $\rho$ together with the channel $\mathcal{E}$, from which the final state $\mathcal{E}(\rho)$ is obtained. In this regard, note that the pair of marginals alone, namely $\rho$ and $\mathcal{E}(\rho)$, is not enough to characterize a trajectory-like description of the process: one also needs an object that keeps track of how input and output are connected by the channel. This is what the two-time operators introduced below are designed to capture.

The reverse process requires more care. As discussed in the introduction, our focus here is not on a reverse protocol specified operationally from the outside, but on an \textit{inferential reverse} constructed from Bayesian retrodiction. In this perspective, the input state of the reverse process, denoted $\tau$, represents the information from which the retrodiction starts. It may coincide with the output of the forward process, but it need not do so; indeed, there are prominent examples in which it should instead be the thermal state \cite{crooks-theorem,tasaki2000jarzynski}, or the decorrelated state of the system and the bath \cite{EspositoNJP}. As for the reverse map, we choose the \textit{Petz transpose map} \cite{petz1,petz}
\begin{align}
    \mathcal{R}_\mathcal{E}^\gamma(\bullet) \coloneqq \sqrt{\gamma}\,\mathcal{E}^\dag\left[\mathcal{E}(\gamma)^{-1/2}\bullet \mathcal{E}(\gamma)^{-1/2}\right]\sqrt{\gamma} \,.
\end{align}
This map has recently been invoked to derive quantum fluctuation theorems \cite{kwon-kim,BS21}. More importantly for our purposes, it is a quantum analog of Bayesian retrodiction \cite{Parzygnat2023axiomsretrodiction,minimum-update-principle}; as such, it depends on a \textit{prior} $\gamma$, which may differ from the actual initial state $\rho$. The usual thermodynamical condition of detailed balance is recovered by choosing as prior a steady state of $\mathcal{E}$. For the derivations below, however, we keep the prior free, while being aware that not every choice is equally relevant from the physical point of view.

The Petz transpose map serves as the Bayesian inverse at the level of quantum channels, rather than at the level of classical probability distributions. This differs from approaches that first translate quantum dynamics into classical data, such as by Bayesian updating of phase-space distributions \cite{bartolotta2016bayesian} or by extracting classical random variables from Bayesian networks \cite{micadei2020quantum}, and subsequently define entropy production on the classical level. 
Our construction avoids the ambiguities in quantum-to-classical coarse-graining and reveals features that may be obscured by translation to classical trajectories. Our approach is more general, in the sense that it reduces to classical Bayesian updating when commutativity relations are satisfied, as shown in \cref{sec:classical_case,sec:tpm}.

\subsection{Choice of quantum objects}

We now turn to the concrete realization of the forward and reverse quantum objects introduced in the previous section. Our aim is to represent, for a process described by a single CPTP map $\mathcal{E}$, the inferential input-output structure that underlies the entropy production operator. In the forward direction, the relevant data of the ``trajectory'' are the input state $\rho$ and the output state $\mathcal{E}(\rho)$.

A first attempt to define a two-time quantum object corresponding to the forward process is
\begin{align} \label{eq:QFtilde}
    \tilde{Q}_F(\rho)&\coloneqq(\one\otimes\sqrt{\rho^T})\,C_\mathcal{E}\,(\one\otimes\sqrt{\rho^T}).
\end{align}
where $C_\mathcal{E}\coloneqq\sum_{i,j} \mathcal{E}(\ketbra{i}{j})\otimes\ketbra{i}{j}$ is the Choi operator
\cite{choi1975completely} for channel $\mathcal{E}$, defined with a basis $\{\ket{i}\}$ of the input system.
The operator $\tilde{Q}_F(\rho)$ is non-negative and unit-trace, hence a valid bipartite state. Moreover, it encodes both the initial and the final states, which can be recovered as $\textrm{Tr}_A[\tilde{Q}_F(\rho)]=\rho^T$ and $\textrm{Tr}_B[\tilde{Q}_F(\rho)]=\mathcal{E}(\rho)$.
Objects of this kind have appeared in studies of quantum conditional probability \cite{leifer2006quantum,chruscinski2020quantum} and observational entropy \cite{bai2023observational}, and are closely related to the Leifer--Spekkens state over time \cite{leifer2007conditional,leifer2013towards}. We refer the reader to Refs.~\cite{horsman2017can,fullwood2022quantum,lie2023uniqueness} for various proposals of states over time and their properties.

We now construct the corresponding object for the reverse process, in the inferential sense discussed above. As shown in~\cite{bai2023observational}, the Choi operator of $\mathcal{R}_\mathcal{E}^\gamma$ can be related to that of $\mathcal{E}$ as follows:
\begin{align*}
    &C_{\mathcal{R}_\mathcal{E}^\gamma}^T\nonumber= (\sqrt{\tau}\mathcal{E}(\gamma)^{-1/2} \otimes\sqrt{\gamma^T})\,C_\mathcal{E}\,(\mathcal{E}(\gamma)^{-1/2}\sqrt{\tau}\otimes\sqrt{\gamma^T})\,.
\end{align*}
The corresponding two-time object is then defined in analogy with \cref{eq:QFtilde} as
\begin{align}
    &\tilde{Q}_R^\gamma(\tau)\coloneqq (\sqrt{\tau}\otimes\one)C_{\mathcal{R}_\mathcal{E}^\gamma}^T (\sqrt{\tau}\otimes\one)\\
    &=(\sqrt{\tau}\mathcal{E}(\gamma)^{-1/2} \otimes\sqrt{\gamma^T})\,C_\mathcal{E}\,(\mathcal{E}(\gamma)^{-1/2}\sqrt{\tau}\otimes\sqrt{\gamma^T})\,.\nonumber
\end{align}
These constructions show that a two-point process can indeed be represented by a single bipartite quantum state, both for the forward process and for its inferential reverse.

However, $\sigma[\tilde{Q}_F,\tilde{Q}_R]$ does not recover the thermodynamical relations that we seek. For this reason, we turn instead to the following ``pseudotransposes'' of the $\tilde{Q}$'s:
\begin{align}
    Q_F(\rho)&\coloneqq\sqrt{C_\mathcal{E}}\,(\one\otimes \rho^T)\,\sqrt{C_\mathcal{E}}\,,\label{QF}\\
    Q_R^\gamma(\tau)&\coloneqq\sqrt{C_\mathcal{E}}\left(\mathcal{E}(\gamma)^{-1/2}\tau \mathcal{E}(\gamma)^{-1/2}\otimes \gamma^T\right)\sqrt{C_\mathcal{E}}\,.\label{QR}
\end{align}
These operators are again valid bipartite states, but they also have a feature that will be important in what follows: their dependence on $\rho$ and $\tau$ is \textit{linear}. In fact, as shown in \cref{app:interpretation_tQ}, $Q_F$ (up to transposition) and $Q_R^\gamma$ are both well-defined \textit{quantum channels} acting on $\rho$ and on $\tau$, respectively. More precisely, $Q_F$ is equivalent to the complementary channel of $\mathcal{E}$, up to a transpose. Similarly, $Q_R^\gamma$ is equivalent to the complementary channel of the reverse map $\mathcal{R}_{\mathcal{E}}^\gamma$.

For the sake of readability, whenever there is no risk of confusion we will omit the explicit dependence on $\rho$ and $\tau$ and simply write $Q_F(\rho)$ and $Q_R^\gamma(\tau)$ as $Q_F$ and $Q_R^\gamma$, respectively.

\subsection{Explicit expressions for the average entropy production}

A key connection with thermodynamics is the explicit expression of the average entropy production.

For channels with full-rank Choi operator, we find (\cref{app:avggen})
\begin{align}\label{avggen}
&\Sigma=\avg{\sigma[Q_F,Q_R^\gamma]}_F\nonumber\\
   &=D_{\textrm{BS}}(\rho\|\gamma)-\textrm{Tr}\left[\mathcal{E}(\rho)\log\left(\mathcal{E}(\gamma)^{-1/2}\tau \mathcal{E}(\gamma)^{-1/2}\right)\right]\,.
\end{align}
Two special cases are worth expanding upon. First, when $[\mathcal{E}(\gamma),\tau]=0$, this takes a more familiar-looking form
\begin{align}
    \Sigma= D_{\textrm{BS}}(\rho\|\gamma)-D(\mathcal{E}(\rho)\|\mathcal{E}(\gamma))+D(\mathcal{E}(\rho)\|\tau)\;,\label{qversion2}
\end{align}
 where $D$ is the Umegaki’s relative entropy defined in \cref{umegaki}. This condition is the same under which the minimum change argument recovers the Petz map \cite{minimum-update-principle}. It is independent of the ``true'' initial and final states, as it involves the posterior of the prior and the state from which one decides to start the reverse process. This expression also makes contact with the early result of
Deffner and Lutz~\cite{DeffnerLutz11}. In the thermodynamic choice
\(\gamma=\rho^{\rm eq}\) and \(\tau=\mathcal E(\rho)\), and when the
relevant operators commute, \cref{qversion2} reduces to the standard open-system
entropy production. We'll see some examples below.

Second, when the reverse process starts from the final state of the forward process, i.e., when $\tau=\mathcal{E}(\rho)$, we can prove the inequality (\cref{app:tau=Erho})
\begin{align}
\Sigma\geq D_{\textrm{BS}}(\rho\|\gamma)&-D(\mathcal{E}(\rho)\|\mathcal{E}(\gamma))\,. \label{eq:tau=Erho}
\end{align}
If further $[\mathcal{E}(\gamma),\mathcal{E}(\rho)]=0$, \cref{eq:tau=Erho} becomes an equality, following from \cref{qversion2}.

Turning to rank-deficient channels, we study only the important case of unitary channels $\mathcal{U}$. In this case, the Choi operator is a rank-one projector, thus $Q_R^\gamma(\tau)=Q_F(\rho)$ holds for all choices of states $\rho$, $\gamma$, and $\tau$. Therefore,
\begin{align}\label{eq:unitary_gives_zero}
    \Sigma& \stackrel{\mathcal{E}=\mathcal{U}}{=} 0\,.
\end{align}
This is consistent with the inferential interpretation of entropy production, since unitary channels preserve information. This informational reversibility of unitary channels is however different from energetic reversibility, where a thermodynamically reversible process may involve external drive and be non-unitary for the system alone.

\subsection{Locality in time and the Heisenberg work operator}\label{sec:locality}

When $\rho$ and the Choi operator $C_\mathcal{E}$ are full rank\footnote{In case $\rho$ or $C_{\mathcal{E}}$ is rank-deficient, one can observe the same by defining $U$ with the polar decomposition of $(\one\otimes\sqrt{\rho^T})\sqrt{C_{\mathcal{E}}}$ as $(\one\otimes\sqrt{\rho^T})\sqrt{C_{\mathcal{E}}}=U\sqrt{Q_F}$.}, we have
\begin{align}\label{locality}
    U\;\sigma[Q_F,Q_R^\gamma]\;U^\dag &= \one\otimes \log \sqrt{\rho^T}(\gamma^T)^{-1}\sqrt{\rho^T}\nonumber\\
    &\quad - \log \mathcal{E}(\gamma)^{-1/2}\tau\mathcal{E}(\gamma)^{-1/2} \otimes \one\;,
\end{align}
where $U:=(\one\otimes\rho^T)^{-1/2}C_{\mathcal{E}}^{-1/2}\sqrt{Q_F}$ is the global unitary such that $U Q_F U^\dag=\tilde{Q}_F$, with $\tilde{Q}_F$ defined in \cref{eq:QFtilde}. \cref{locality} shows that, after the global change of basis implemented by $U$, the entropy production operator decomposes into two commuting local terms, one acting only on the input system and the other only on the output system. This is highly non-trivial: the spectrum is then constrained by only two local spectra of size at most $d$, whereas a generic bipartite self-adjoint operator on a $d^2$-dimensional space may have $d^2$ distinct eigenvalues. We call this property ``locality in time''.

This terminology is motivated by the classical case. There, when the fluctuating entropy production can be written as the sum of one term depending only on the initial state and one term depending only on the final state, the corresponding reverse statistics are necessarily of Bayesian form~\cite{AwBS}. In this sense, locality in time is the operator counterpart of the retrodictive structure identified in the classical theory of fluctuation relations.

If moreover $[(\one\otimes\rho^T),C_{\mathcal{E}}]=0$ is satisfied (notably, for classical-quantum channels), then $U=\one$ and locality in time holds exactly, at the operator level. Otherwise, locality in time does not hold for the operator $\sigma$ itself; nevertheless, because \cref{locality} is a unitary equivalence, the eigenvalues of $\sigma$ still split into sums of an initial-time contribution and a final-time contribution.

To show that the objects discussed here acquire physical meaning in a familiar setting, consider the special case, in which $\mE$ is unital, the prior $\gamma$ is uniform (hence a steady state), the initial state of the system $\rho$ is the Gibbs state with respect to the initial Hamiltonian $H$, and the initial state of the reverse process $\tau$ is the Gibbs state with respect to the final Hamiltonian $H'$. In this case, \cref{locality} reduces to the familiar expression
\begin{align}
U\;\sigma[Q_F,Q_R^\gamma]\;U^\dag &= \beta(\one\otimes H'-H\otimes\one-\Delta F)\\
    &= \beta(\Omega-\Delta F)\;,
\end{align}
where $\Omega$ is the so-called \textit{Heisenberg operator of work}~\cite{heisenberg-work-op,Hovhannisyan2024energyconservation,rubino2024revisingquantumworkfluctuation}, but here generalized beyond the case of a unitary channel. We leave the exploration of further connections between our entropy production operator and the theory of work operators to future investigations.

\subsection{Channel composition}

We now ask how entropy production behaves under the concatenation of two channels. Let $\mathcal{E}_1$ and $\mathcal{E}_2$ be composable channels, and write their composition as
\[
\mE \coloneqq \mathcal{E}_2\circ\mathcal{E}_1 \;.
\]
A typical example is a channel followed by a measurement, $(\mathcal{E}_1,\mathcal{E}_2)=(\mathcal{E},\mathcal{M})$.

Let the initial state and prior be $\rho$ and $\gamma$, and let $\tau_1$ and $\tau_2$ denote the new information available after the first and second steps, respectively. The reverse of the second step is then constructed with respect to the prior $\mathcal{E}_1(\gamma)$.

The two-time operators for the total process $\mE=\mathcal{E}_2\circ\mathcal{E}_1$ are
\begin{align*}
    Q_F(\rho)
    &= \sqrt{C_{\mE}}(\one\otimes\rho^T)\sqrt{C_{\mE}} \\
    Q_R^\gamma(\tau_2)
    &= \sqrt{C_{\mE}}
       \left(\mE(\gamma)^{-1/2}\tau_2\mE(\gamma)^{-1/2}\otimes\gamma^T\right)
       \sqrt{C_{\mE}} \;.
\end{align*}
For the two steps separately, the corresponding operators are
\begin{align*}
    Q_{F,1}(\rho)
    &= \sqrt{C_{\mathcal{E}_1}}(\one\otimes\rho^T)\sqrt{C_{\mathcal{E}_1}} \\
    Q_{R,1}^\gamma(\tau_1)
    &= \sqrt{C_{\mathcal{E}_1}}
       \left(\mathcal{E}_1(\gamma)^{-1/2}\tau_1\mathcal{E}_1(\gamma)^{-1/2}\otimes\gamma^T\right)
       \sqrt{C_{\mathcal{E}_1}} \;,
\end{align*}
and
\begin{align*}
    Q_{F,2}(\mathcal{E}_1(\rho))
    &= \sqrt{C_{\mathcal{E}_2}}
       \left(\one\otimes\mathcal{E}_1(\rho)^T\right)
       \sqrt{C_{\mathcal{E}_2}} \\
    Q_{R,2}^{\mathcal{E}_1(\gamma)}(\tau_2)
    &= \sqrt{C_{\mathcal{E}_2}}
       \left(\mE(\gamma)^{-1/2}\tau_2\mE(\gamma)^{-1/2}\otimes\mathcal{E}_1(\gamma)^T\right)
       \sqrt{C_{\mathcal{E}_2}} \;.
\end{align*}

The associated entropy production operators are
\begin{align*}
    \sigma_1
    &\coloneqq \sigma[Q_{F,1}(\rho),Q_{R,1}^\gamma(\tau_1)] \\
    \sigma_2
    &\coloneqq \sigma[Q_{F,2}(\mathcal{E}_1(\rho)),Q_{R,2}^{\mathcal{E}_1(\gamma)}(\tau_2)] \\
    \sigma_{12}
    &\coloneqq \sigma[Q_F(\rho),Q_R^\gamma(\tau_2)] \;,
\end{align*}
with corresponding averages
\begin{align*}
    \Sigma_1
    &\coloneqq \Tr[Q_{F,1}(\rho)\,\sigma_1] \\
    \Sigma_2
    &\coloneqq \Tr[Q_{F,2}(\mathcal{E}_1(\rho))\,\sigma_2] \\
    \Sigma_{12}
    &\coloneqq \Tr[Q_F(\rho)\,\sigma_{12}] \;.
\end{align*}

Then, when the quantities are well-defined, we prove in \cref{app:superadditivity} that
\begin{align}
    \Sigma_1+\Sigma_2 \geq \Sigma_{12} \;.
    \label{eq:superadditive}
\end{align}
Thus the average entropy production is, in general, not additive under channel composition, but the sum of the stepwise contributions is always no smaller than the entropy production of the overall process.

The equality condition also has a natural interpretation. Equality in \cref{eq:superadditive} holds if and only if
\[
\tau_1=\mathcal{E}_1(\rho)
\qquad\text{and}\qquad
[\mathcal{E}_1(\rho),\mathcal{E}_1(\gamma)]=0 \;.
\]
That is, there is no excess contribution at the intermediate time precisely when the new information at the intermediate step coincides with the actual forward output (i.e., $\tau_1=\mathcal{E}_1(\rho)$), and this commutes with the expected intermediate state (i.e., $[\mathcal{E}_1(\rho),\mathcal{E}_1(\gamma)]=0$).

It is then natural to introduce the non-negative quantity
\[
    \avg{\Upsilon_{1:2}}_F \coloneqq \Sigma_1+\Sigma_2-\Sigma_{12} \ge 0 \;.
\]
This may be viewed as a correlation-like excess entropy production associated with the intermediate time slice. Whether this analogy can be promoted to a genuine information-theoretic interpretation, akin to mutual information, is left for future investigation.

\section{Case studies}

In this section, we illustrate the general framework developed above through a number of representative classes of quantum channels. The purpose of these examples is twofold. First, they show how the abstract construction of the entropy-production operator can be evaluated explicitly in concrete settings. Second, they help clarify which features of our proposal reproduce familiar thermodynamic expectations and which ones instead reveal genuinely quantum departures from the classical picture.

We begin with Gibbs channels, which provide the most natural benchmark from the viewpoint of quantum thermodynamics. We then turn to other special classes of channels, chosen so as to highlight different structural aspects of the framework, such as the role of Bayesian retrodiction, the effect of non-commutativity, and the relation with more classical notions of irreversibility. Together, these case studies provide a more concrete understanding of both the scope and the limitations of our construction.

\subsection{Gibbs channels}
\label{sec:cases}

As we said in the introduction, it is widely accepted that the average entropy production for Gibbs channels is given by \cref{eq2}. Our proposed expressions do not always recover that result. In what follows we numerically investigate this discrepancy.

Let us first set $\gamma=\mathcal{E}(\gamma)=\rho^{\textrm{eq}}$. If the reverse process is chosen to start from $\tau=\rho^{\textrm{eq}}$ as well, then we are in a special case of \cref{qversion2}, and therefore $\Sigma=D_{\textrm{BS}}(\rho\|\gamma)$. If instead the reverse process starts where the forward one ends, namely $\tau=\mathcal{E}(\rho)$, then we are in a special case of \cref{eq:tau=Erho}. In that case, our expression is generically larger than the conventional formula \cref{eq2}. Equality is recovered if both $[\rho,\rho^{\textrm{eq}}]=0$, so that the Belavkin--Staszewski relative entropy reduces to the Umegaki one, and $[\mathcal{E}(\rho),\rho^{\textrm{eq}}]=0$, so that equality holds in \cref{eq:tau=Erho}. These are nontrivial restrictions. In particular, they do not imply $[\rho,\mathcal{E}(\rho)]=0$, nor do they force the channel $\mathcal{E}$ to be measure-and-prepare.

To gain some intuition about the size and structure of this difference, we consider a one-parameter family of collisional models for a single qubit that describe thermalization \cite{scarani2002} or, more generally, homogenization \cite{ziman02}. A special case of these models was also used as an example in Ref.~\cite{twesh}. In these models, a qubit undergoes $n$ iterations of the same channel $\mathcal{N}$. The channel is defined by preparing an ancilla qubit in a reference state $\xi$ and letting it interact with the system through an interaction that preserves the number of excitations in the eigenbasis of $\xi$, denoted $\{\ket{0},\ket{1}\}$. Under these conditions, $\xi$ is the fixed point of the channel, and $\mathcal{N}^n(\rho)\rightarrow \xi$ for all $\rho$ in the limit of large $n$.

To keep the number of parameters limited, we focus on the most symmetric member of this family, namely the one generated by the interaction $H_{\textrm{int}}\propto XX+YY+ZZ$. This is the unique interaction that preserves the number of excitations in every basis, and it generates the unitary
\begin{align*}
U=\cos\phi\,\mathbb{I}+i(\sin\phi) \,U_{\textrm{swap}}\;.    
\end{align*}
Note that the case $n=1$ coincides with the example studied in Ref.~\cite{twesh}, motivated there by the conservation of the non-commuting ``charges'' defined by the Pauli matrices.

The dynamics of the system qubit after $n$ collisions is given by
\begin{equation}
\begin{aligned}
    \sandwich{0}{\mathcal{N}^n(\rho)}{0}&=c^{2n}\sandwich{0}{\rho}{0}+(1-c^{2n})\sandwich{0}{\xi}{0}\,,\\
    \sandwich{0}{\mathcal{N}^n(\rho)}{1}&=k^n\,\sandwich{0}{\rho}{1} 
\end{aligned}\label{eq:channel_N}
\end{equation}
where $c=\cos \phi$ and
\begin{align*}
    k=\frac{1}{2}\Big[1+\cos 2\phi+i\sin 2\phi\big(2\sandwich{0}{\xi}{0}-1\big)\Big]\;.
\end{align*}
For comparison with Refs.~\cite{scarani2002,ziman02}, we denote here by $k$ what was written there as $\lambda\cos\phi$.

We compute the average entropy production $\Sigma$ for the family of channels $\mathcal{N}^n$, for varying $n$. These channels are full rank as long as $\xi$ is not pure, so \cref{avggen} applies. In the figures below, we represent the $(x,z)$ plane of the Bloch sphere of $\rho$, for given choices of $\gamma$ and $\tau$, and use a heatmap to show the value of $\Sigma$.

\begin{figure}
    \centering
    \includegraphics[width=1\linewidth]{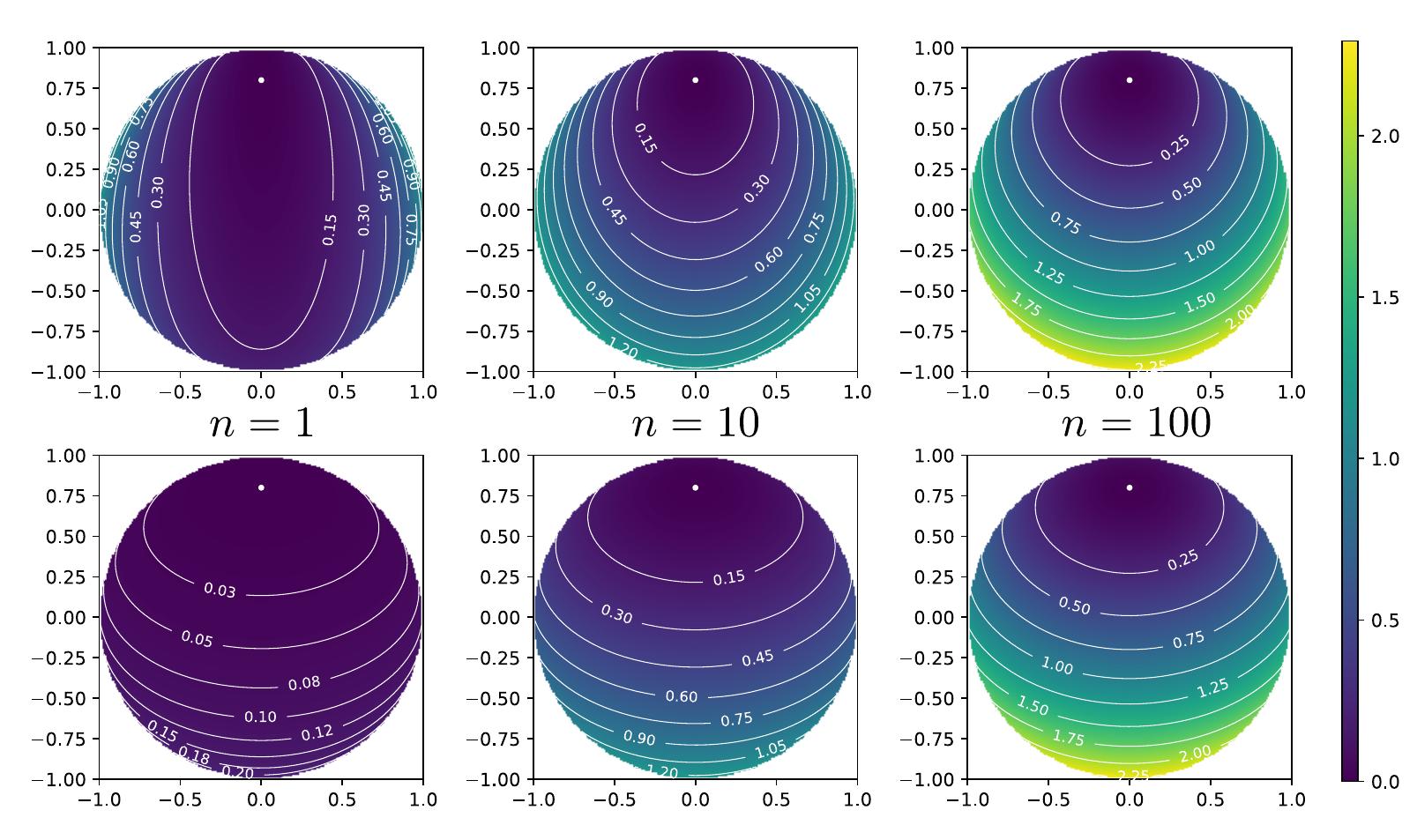}
    \caption{Average entropy production for $\gamma=\xi=\mathcal{N}^n(\gamma)$ (the dot in the figure) and $\tau=\mathcal{N}^n(\rho)$. For the numerics, $\sandwich{0}{\xi}{0}=0.9$, $\phi=0.2$, and $n$ given in the figure. Top row: average entropy production as given in \cref{avggen}. Bottom row: average entropy production as given by \cref{eq2}. The two formulas agree when $[\tau,\mathcal{N}^n(\gamma)]=0$ and $[\rho,\gamma]=0$ hold, which is true for all $n$ on the $z$ axis.
    }
    \label{fig:erho}
\end{figure}

Our first case study, shown in Figure~\ref{fig:erho}, corresponds to the most standard thermodynamic choice: the prior $\gamma$ is taken to be the fixed point $\xi$ of the channel, which plays the role of the Gibbs state $\rho^{\textrm{eq}}$, and the reverse process starts where the forward one ends, namely $\tau=\mathcal{N}^n(\rho)$. The top row shows our prediction for three values of $n$, while the bottom row shows the more conventional expression $D(\rho\|\xi)-D(\mathcal{N}^n(\rho)\|\xi)$ from \cref{eq2}. As anticipated by \cref{eq:tau=Erho}, our expression is generically slightly larger.

The two formulas coincide for all $n$ when $\rho$ lies on the $z$ axis of the Bloch sphere, that is, when $\rho$ is diagonal in the basis $\{\ket{0},\ket{1}\}$. Indeed, in that case $\mathcal{N}^n(\rho)$ remains diagonal as well, so $[\tau,\mathcal{N}^n(\gamma)]=[\mathcal{N}^n(\rho),\xi]=0$, and the entropy production is given by \cref{qversion2}. In that expression, the last term vanishes, and $D_{\textrm{BS}}(\rho\|\gamma)=D(\rho\|\gamma)$ because $[\rho,\gamma]=0$. In the limit $n\rightarrow\infty$, we also recover equality in \cref{eq:tau=Erho}, because then $\tau=\mathcal{N}^n(\rho)\rightarrow \xi$. For the parameter range shown in the figure, we numerically observe that the entropy production does not decrease with the number of collisions, for any initial state $\rho$ in the plotted plane.

\begin{figure}
    \centering
    \includegraphics[width=1\linewidth]{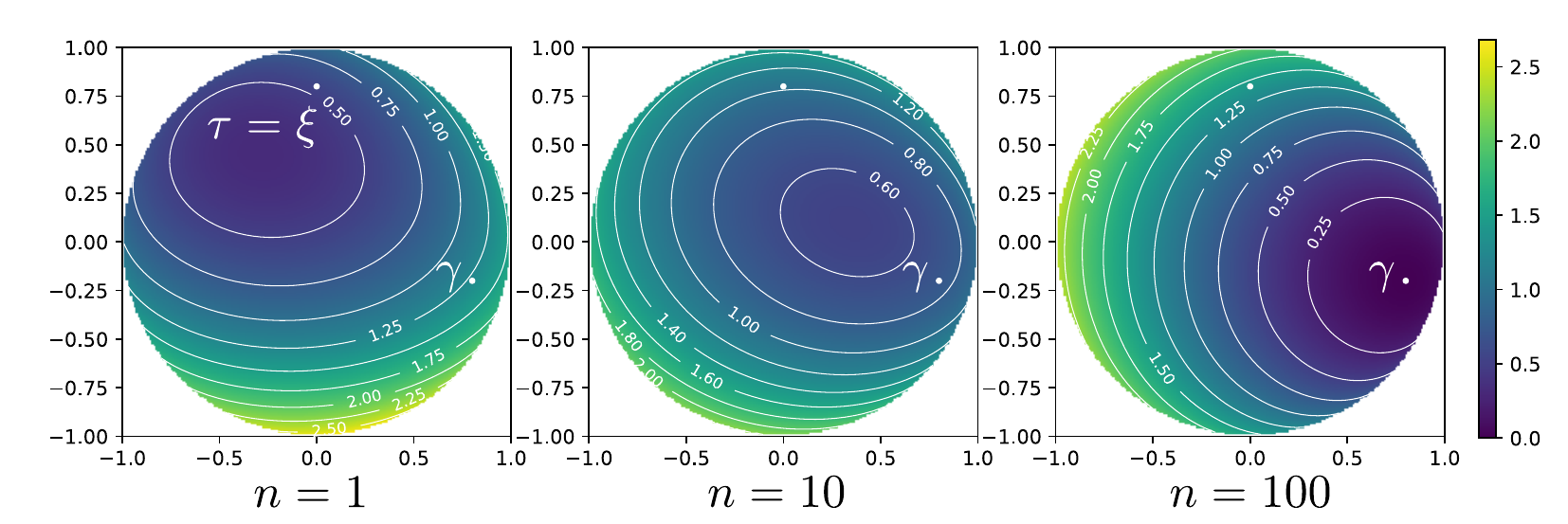}
    \caption{Average entropy production as defined in \cref{avggen}, for $\tau=\xi$ and $\gamma$ diagonal in a different basis. For the numerics, $\sandwich{0}{\xi}{0}=0.9$, $\phi=0.2$, and $n$ as indicated in the figure. The Bloch vector of $\gamma$ is $(0.8,0,-0.2)$. As the number $n$ of collisions increases, the entropy production tends to $D_{\textrm{BS}}(\rho\|\gamma)$, as clearly shown by the level curves in the right panel, which center around $\gamma$ and become less and less dependent on $\tau$.} 
    \label{fig:tauxi}
\end{figure}

As a second case study, shown in Figure~\ref{fig:tauxi}, we fix $\tau=\xi$ but choose a prior $\gamma$ that is diagonal in a different basis. This choice is not the standard thermodynamic one, but it is useful for isolating the inferential role of the prior in our framework. For large $n$, we have $\mathcal{N}^n(\rho)\approx\mathcal{N}^n(\gamma)\approx \tau$, so the second term of the entropy production in \cref{avggen} tends to zero, and the result is dominated by $D_{\textrm{BS}}(\rho\|\gamma)$. In particular, one sees that for some $\rho$ in a neighborhood of $\gamma$, the entropy production \textit{decreases} as the number of collisions increases. This behavior is not specifically quantum: it can already occur in the classical case, for instance for a suitable choice of the reverse starting point combined with a permutation.

\subsection{Quantum-classical channels, and irreversibility of measurement}\label{sec:qc-channels}

We now consider quantum-classical channels, namely channels with fully quantum input and commutative output. These channels naturally describe measurements performed on the input system, with the outcome probabilities encoded in the diagonal elements of the output state. From this perspective, the corresponding entropy production can be interpreted as a form of irreversibility associated with measurement \cite{elouard2017role,safranek2019a,safranek2019b,safranek2021brief,buscemi2022observational}. 
Formally, we consider channels of the form
\begin{align}
    \mathcal{M}(\rho) = \sum_i \Tr[\Pi_i \rho] \ketbrasame{i}\;, \label{eq:measurement_channel}
\end{align}
where $\{\Pi_i\}_i$ is a POVM and $\{\ket{i}\}_i$ is an orthonormal basis. By construction, both $\mathcal{M}(\rho)$ and $\mathcal{M}(\gamma)$ are diagonal in the basis $\{\ket{i}\}$. It is then natural to assume that the starting point of the reverse process, $\tau$, is diagonal in the same basis.

This commutative structure simplifies the operator $\sigma[Q_F,Q_R^\gamma]$ considerably. In the quantum-classical case, the operators $C_{\mathcal{M}}$, $\mathcal{M}(\rho)\otimes\one$, $\mathcal{M}(\gamma)\otimes\one$, and $\tau\otimes\one$ all commute. As a result, the entropy production operator takes the block-diagonal form
\begin{align}
    \sigma[Q_F,Q_R^\gamma]=& \sum_i \ketbrasame{i}\otimes \sigma\left[\sqrt{\Pi_i}\rho\sqrt{\Pi_i},\sqrt{\Pi_i}\gamma\sqrt{\Pi_i}\right]^T \nonumber \\
    &\mathrel{\phantom{=}}-(\ln\tau-\ln\mathcal{M}(\gamma))\otimes\one\;,
\end{align}
where
\begin{align}
    &\sigma \left[\sqrt{\Pi_i}\rho\sqrt{\Pi_i},\sqrt{\Pi_i}\gamma\sqrt{\Pi_i}\right] \nonumber\\
    &= \log\sqrt{\sqrt{\Pi_i}\rho\sqrt{\Pi_i}}\Big(\sqrt{\Pi_i}\gamma\sqrt{\Pi_i}\Big)^{-1}\sqrt{\sqrt{\Pi_i}\rho\sqrt{\Pi_i}}\;.
\end{align}
Thus, the quantum contribution is organized into blocks labeled by the classical measurement outcomes, while the output-side contribution is purely classical.

At the level of the average, \cref{avggen} reduces to
\begin{align}
\Sigma=D_{\textrm{BS}}(\rho\|\gamma)&-D(\mathcal{M}(\rho)\|\mathcal{M}(\gamma))+D(\mathcal{M}(\rho)\|\tau)\,, \label{eq:DBS-D+D2}
\end{align}
where the last two terms are relative entropies of classical distributions, namely Kullback-Leibler divergences. This expression makes transparent the hybrid nature of the quantum-classical case: the input-side contribution remains genuinely quantum, while the output-side contribution is entirely classical.

This structure also connects naturally with the framework of \textit{observational entropy}~\cite{safranek2019a,safranek2019b,safranek2021brief,buscemi2022observational}. In that setting, the entropy associated with a measurement $\mathsf M$ is interpreted as the sum of the intrinsic uncertainty of the state and the additional uncertainty induced by the observation, which need not be maximally informative. In its version with an arbitrary quantum prior $\gamma$, observational entropy is given by
\begin{align}
    S_{\mathsf M}(\rho) = S(\rho)+D_{\textrm{BS}}(\rho\|\gamma)-D(\mathcal{M}(\rho)\|\mathcal{M}(\gamma))\;.
\end{align}
If, following the observational-entropy narrative, one identifies the new information obtained from the measurement with the actual measurement statistics, namely $\tau=\mathcal{M}(\rho)$, then the last term in \cref{eq:DBS-D+D2} vanishes. In that case, our average entropy production becomes
\begin{align}
    \Sigma=S_{\mathsf M}(\rho)-S(\rho)\;.
\end{align}
Therefore, for quantum-classical channels, the average entropy production associated with our inferential construction coincides exactly with the entropy increase induced by the observation. This provides a clear operational meaning for $\Sigma$ in the context of measurement, and shows that the present framework naturally extends previously studied notions of entropy associated with coarse-grained observation.

\subsection{Classical case} \label{sec:classical_case}

We now discuss the classical limit of our construction. This limit is recovered when all the relevant states are diagonal in a common basis, and the channel is a measure-and-prepare channel on that same basis. In that situation, the quantum problem reduces to an ordinary stochastic process, and one should recover the usual classical notion of entropy production. This is indeed what happens.

Under these assumptions, the operators $C_{\mathcal{E}}$, $\one\otimes\rho^T$, $\one\otimes\gamma^T$, $\mathcal{E}(\rho)\otimes\one$, $\mathcal{E}(\gamma)\otimes\one$, and $\tau\otimes\one$ all commute. Therefore, the entropy production operator becomes diagonal and takes the form
\begin{align}\label{Sop_clax}
    &\sigma[Q_F,Q_R^\gamma] \nonumber \\
    &= \one\otimes(\log\rho-\log\gamma) - \mathopen{\big(}\log\tau-\log\mathcal{E}(\gamma)\big)\otimes\one\\
    &=\sum_{i,j}\log\frac{p_i\,\lambda_j }{\pi_i\, q_j}\,\ketbrasame{j}\otimes\ketbrasame{i}
\end{align}
where $p$, $\pi$, $\lambda=\varphi\pi$ ($\varphi$ is the stochastic matrix induced by $\mathcal{E}$ on the chosen basis), and $q$ are the eigenvalues of $\rho$, $\gamma$, $\mathcal{E}(\gamma)$, and $\tau$, respectively. As expected, $\sigma[Q_F,Q_R^\gamma]$ is diagonal, and its eigenvalues are exactly the classical values of entropy production associated with the corresponding input-output pairs.

At the level of averages, \cref{avggen} reduces to the corresponding classical expression. Indeed, in the present commutative setting one has $D_{\textrm{BS}}(\rho\|\gamma)=D(\rho\|\gamma)$, and all Umegaki relative entropies reduce to Kullback-Leibler divergences for the corresponding probability distributions. Thus, our construction recovers the standard classical log-ratio structure both at the operator level and at the level of the average entropy production. Moreover, this conclusion continues to hold even for non-full-rank channels, by continuity of Umegaki relative entropy \cite{audenaert2005,schindler2023continuity,bluhm2023general}.

It is worth stressing that the classical case is recovered by measure-and-prepare channels, not by unitary channels that merely admit a classical interpretation, such as the identity channel or a permutation of basis elements. The reason is that unitary channels do not define a genuine joint input-output probability distribution. In our framework, they instead fall under the general statement already noted in \cref{eq:unitary_gives_zero}, namely that $\Sigma=0$ for unitary channels. In this sense, classical stochasticity is recovered not from reversible relabelings, but from channels that genuinely realize a classical input-output process.

This distinction also explains why the classical case allows non-zero entropy production for reversible processes when $\tau\neq\mathcal{E}(\rho)$, whereas the quantum unitary formula [\cref{eq:unitary_gives_zero}] gives $\Sigma=0$ unconditionally. In the quantum case, the output of a unitary channel is uniquely $\mathcal{E}(\rho)$. The only way to obtain a different state $\tau$ from which to start the reverse process is to have performed a measurement on the output, thereby extracting partial information and collapsing the state. Therefore, the case $\tau\neq\mathcal{E}(\rho)$ indeed involves implicitly a composite process of a unitary followed by a measurement, and thus the unitary formula is not applicable. In the classical setting, states are already diagonal in the measurement basis, and thus the measurement is naturally absorbed.

\subsection{Two-point measurement scheme}\label{sec:tpm}

We conclude with the two-point measurement scheme, which was the first framework in which quantum fluctuation theorems were formulated \cite{tasaki2000jarzynski,kurchan}. We present it in the generalized form introduced in Ref.~\cite{BS21}, then show the Tasaki-Crooks relation as a special case.

A classical label $i$, distributed according to probabilities $p_i$, is used to prepare a state $\rho_i$. Keeping track of the label in an explicit register, the input state is
\begin{align}\label{rhotas}
    \rho=\sum_i p_i\,\rho_i\otimes \ketbrasame{i}\;.
\end{align}
The prior has the same form,
\begin{align}
    \gamma=\sum_i \pi_i\,\rho_i\otimes \ketbrasame{i}\;,
\end{align}
so that $[\rho,\gamma]=0$.

After the subsequent evolution, a measurement is performed on the system. Absorbing the evolution into the definition of the states $\rho_i$, one obtains
\begin{align}\label{Erhotas}
    \mathcal{E}(\rho)=\sum_i p_i\,\mathcal{M}(\rho_i)\otimes\ketbrasame{i}
    =\sum_{i,j} p_i \tr{\rho_i\Pi_j} \ketbrasame{j}\otimes\ketbrasame{i}\;,
\end{align}
which defines an effectively classical process with transition probabilities
\begin{align}\label{phitas}
    \varphi(j|i)=\tr{\rho_i\Pi_j}\;.
\end{align}
It is then natural to choose the starting point of the reverse process in diagonal form,
\begin{align}\label{tautas}
    \tau=\sum_j q_j\ketbrasame{j}\;.
\end{align}
Next we show how this formalism recovers the Tasaki-Crooks relation \cite{tasaki2000jarzynski}.

The process considered by Tasaki is as follows. The first preparation is made by measuring the energy at time $t_0$ on an initial state $\rho_0$. The Hamiltonian and its eigenstates are denoted by $H_0\ket{\phi_0^i}=E_0^i\ket{\phi_0^i}$, and of course state $\ket{\phi_0^i}$ will be found with probability $p_i=\bra{\phi_0^i}\,\rho_0\ket{\phi_0^i}$. Tasaki assumes from the start that $\rho_0$ is the Gibbs state for $H_0$, whence $p_i=e^{-\beta E_0^i}/Z_0$. Then the state undergoes a unitary evolution, and at later time $t_1$ another energy measurement is performed; the Hamiltonian may have been driven to another one, so that now we have $H_1\ket{\phi_1^j}=E_1^j\ket{\phi_1^j}$. Finally, in order to derive a Crooks-like theorem, Tasaki considers a reverse process that starts from the Gibbs state of $H_1$ and evolves according to $U^{-1}$. 

This physics is captured in the formulas just above as follows: recalling that we absorb the evolution in the definition of the $\rho_i$, \cref{rhotas} becomes
\begin{align}
    \rho=\sum_i p_i\,U\ket{\phi_0^i}\bra{\phi_0^i}U^\dagger\otimes \ketbrasame{i}
\end{align}
and \cref{Erhotas} becomes $\mathcal{E}(\rho)=\sum_{i,j} p_{ij}\, \ketbrasame{i}\otimes \ketbrasame{j}$ with
\begin{align}
    p_{ij}=p_i|\bra{\phi_1^j}\,U\ket{\phi_0^i}|^2=\frac{e^{-\beta E_0^i}}{Z_0}|\bra{\phi_1^j}\,U\ket{\phi_0^i}|^2
\end{align}
which is exactly Tasaki's ``crucial quantity'' (2.4). In particular we identify $\varphi(j|i)=|\bra{\phi_1^j}\,U\ket{\phi_0^i}|^2$ in \cref{phitas}. The initial state of the reverse process is \cref{tautas} with $q_j=\frac{e^{-\beta E_1^j}}{Z_1}$.

Our formalism has an extra element: the prior $\gamma$. Tasaki's formulation is recovered by setting all the $\pi_i$ equal to $\frac{1}{d}$ where $d$ is the number of energy levels of the system. With this, $\lambda_j=(\varphi\pi)_j=\sum_i\varphi(j|i)(1/d)=1/d$ as well, and the entropy production operator \cref{Sop_clax} becomes
\begin{align}
    \sigma[Q_F,Q_R^\gamma]&=\sum_{i,j}\log\left(e^{-\beta(E_0^i-E_1^j)}\frac{Z_1}{Z_0}\right)\,\ketbrasame{j}\otimes\ketbrasame{i}\nonumber\\
    &=\sum_{i,j}\big(\beta(\Delta E_{ji}-\Delta F)\big)\,\ketbrasame{j}\otimes\ketbrasame{i}
\end{align} with $\Delta E_{ji}=E_1^j-E_0^i$ and $\beta F=-\log Z$ as usual. This recovers the expected entropy production on each trajectory, starting and finishing in an energy eigenstate.

\section{Conclusion}
\label{sec:conclusion}

We have developed a fully quantum approach to entropy production by extending the log-ratio structure of stochastic thermodynamics to the non-commutative setting. At the general level, we identified a Hermitian quantum analog of the classical entropy-production variable that preserves the main formal features of the classical construction, including a non-negative average and exact fluctuation relations. We then specialized this framework to quantum channels, constructing the reverse process through Bayesian retrodiction and thereby giving the general operator a concrete inferential interpretation.

This led us to distinguish two related levels of description. At the level of averages, one can write quantum counterparts of classical entropy-production formulas in terms of the states that characterize the process: the actual initial state, the prior state used to define the reverse map, the actual final state, and the state from which the reverse process is taken to start. But such average formulas do not identify \textit{what} fluctuates and is being averaged. The main result of this paper is therefore the construction of the entropy-production operator
\[
\sigma[Q_F,Q_R]\coloneqq
 \log \left\{\sqrt{Q_F}\,Q_R^{-1}\,\sqrt{Q_F}\right\},
\]
built from quantum objects representing the forward and reverse input-output processes. This operator is Hermitian, its expectation value is non-negative, and it satisfies exact integral and detailed fluctuation theorems. It thus preserves the main formal features of the classical log-ratio structure in the non-commutative setting.

For the case of a single CPTP map, we then provided explicit expressions by defining the reverse process through the Petz transpose map with respect to a prior state. In this setting, the framework recovers the classical formula in the commutative limit, yields a natural notion of inferential entropy production for quantum channels, and displays further structural properties, such as locality in time up to unitary equivalence and superadditivity under channel composition. It also admits a clear interpretation for some classes of channels, in particular quantum-classical ones, where the average entropy production is directly related to the entropy increase caused by observation.

At the same time, our results make clear that the fully quantum problem is not a straightforward translation of the classical one. Most notably, our construction does not in general reproduce the standard expression
\[
D(\rho\|\rho^{\textrm{eq}})-D(\mathcal{E}(\rho)\|\rho^{\textrm{eq}})
\]
for Gibbs channels. This mismatch should not be viewed merely as a defect of the present proposal. It may indicate that no single operator-valued notion of entropy production can simultaneously retain all the desiderata that look natural from the classical viewpoint: positivity of the average, exact fluctuation relations, a trajectory-like interpretation, and agreement with the conventional thermodynamic formula for all relevant quantum processes. Of course, it may also be that a more clever construction exists. But it is equally plausible that the obstruction is conceptual rather than technical, and that in the genuinely quantum regime one must relax some assumption that is usually taken for granted, perhaps even the idea that entropy production should always descend from a difference of von Neumann entropies.

This point is related to a broader distinction that deserves emphasis. The present framework quantifies irreversibility as a feature of information and inference: it compares what is predicted in the forward description with what can be retrodicted in the reverse one. In that sense, it belongs primarily to the thermodynamics of information. Energy dissipation enters only indirectly, through physically motivated choices of prior, reverse state, and channel. This helps explain why unitary channels give $\Sigma=0$: from the inferential point of view, no information is lost. By contrast, the usual entropy production formula for Gibbs processes is tied to energetic relaxation toward equilibrium and to the special role of thermal reference states. The tension between these two viewpoints, informational and energetic, is not a bug of the formalism. It may reflect a genuine ambiguity in what one wants ``entropy production'' to mean in quantum theory.

Several open problems follow from this work. First, one may ask whether there exists a different entropy-production operator whose average always reproduces the most familiar average formula. A natural clue comes from the fact that the Petz map coincides with the Bayesian retrodiction map derived from a minimum-change principle~\cite{minimum-update-principle} precisely when $[\mathcal{E}(\gamma),\tau]=0$, which is also the condition under which our average entropy production takes its most familiar form. This suggests replacing \cref{QR} with
\begin{align}
&Q_R^\gamma(\tau)\label{QRvar}\\
&=\sqrt{C_\mathcal{E}}\,\left(\left[\sqrt{\tau}(\sqrt{\tau}\mathcal{E}(\gamma)\sqrt{\tau})^{-1/2}\sqrt{\tau}\right]^{2}\otimes \gamma^T\right )\,\sqrt{C_\mathcal{E}}\,.\nonumber
\end{align}
With this substitution, only the explicit formula \eqref{avggen} would change, while the structural results established from the general operator construction would remain the same, including the formulas recovered when $[\mathcal{E}(\gamma),\tau]=0$.

A second open question concerns locality in time. In the classical theory, the decomposition of entropy production into an initial-time term and a final-time term is tightly connected with Bayesian retrodiction~\cite{AwBS}. In our framework, this property survives at the level of averages, and in the operator setting it survives up to a unitary equivalence. One may still wonder whether there exists a more fundamental construction in which locality in time holds exactly at the operator level for arbitrary quantum processes, not only in special commutative cases. A candidate would be to take the right-hand side of \cref{locality} itself as the entropy-production operator. This choice has attractive formal features, but we have not been able to derive it directly from quantum input-output process objects.

More generally, our results suggest that a fully quantum notion of entropy production may require trade-offs. Other approaches keep a closer trajectory-like analogy, but allow quasi-probabilities or even complex values~\cite{kwon-kim,twesh}. Our construction keeps entropy production real and defined from genuine quantum states, but does not in general recover the standard Gibbs-channel formula and locality in time at the operator level. This suggests that no single definition may capture all aspects of quantum irreversibility equally well.

\section*{Acknowledgements}

We thank Alvaro Alhambra, Felix Binder, \'Angela Capel, Hao-Chung Cheng, Karen Hovhannisyan, Gabriel Landi, Matteo Lostaglio, Eric Lutz, Mark Mitchinson, Giulia Rubino, Philip Strasberg, and Twesh Upadhyaya  for discussions and feedback. We also thank the two anonymous referees whose comments and criticisms helped us reshape these results.

This work is supported by the National Research Foundation, Singapore and A*STAR under its CQT Bridging Grant; by the National Research Foundation, Singapore through the National Quantum Office, hosted in A*STAR, under its Centre for Quantum Technologies Funding Initiative (S24Q2d0009); and by the Ministry of Education, Singapore, under the Tier 2 Grant ``Bayesian approach to irreversibility'' (Grant No.~MOE-T2EP50123-0002).
G.\ B.\ acknowledges support from Guangdong Basic and Applied Basic Research Foundation (Grant No.~2026A1515030035) and the Start-up Fund (Grant No.~G0101000274) from The Hong Kong University of Science and Technology (Guangzhou).
F.~B. acknowledges support from MEXT Quantum Leap Flagship Program (MEXT QLEAP) Grant No.~JPMXS0120319794, and from JSPS KAKENHI, Grants Nos.~23K03230 and 26K00621.

\section*{Data availability statement}

All data that support the findings of this study are included within the article (and any supplementary files).

\bibliography{refsretro}

\onecolumngrid\appendix
\crefalias{section}{appendix}
\crefalias{subsection}{subappendix}

\section{Proof of the detailed fluctuation theorem}\label{app:crooks}

Here we prove \cref{crooksc} with the entropy production defined in \cref{eq:epfwd,eq:eprev}.

Assume $Q_F$ and $Q_R$ to be full rank. Let the singular value decomposition of $\sqrt{Q_F}\,Q_R^{-1/2}$ be
\begin{align}
    \sqrt{Q_F}\,Q_R^{-1/2} = UDV^\dagger\,, \label{eq:singular-decomp}
\end{align}
where $U$ and $V$ are unitary operators, and $D$ is a positive definite diagonal operator. They can be written as
\begin{align}
    U = \sum_k \ketbra{f_k}{k}, \quad V = \sum_k \ketbra{r_k}{k}, \quad D = \sum_k d_k\ketbra{k}{k} , \quad d_k>0 \label{eq:UVS}
\end{align}
for some orthonormal basis $\{\ket{k}\}$. $\{\ket{f_k}\}$ and $\{\ket{r_k}\}$  are the eigenvectors of $\sigma[Q_F,Q_R]$ and $\sigma[Q_R,Q_F]$, which can be seen from
\begin{align}
    e^{\sigma[Q_F,Q_R]} &= \sqrt{Q_F}Q_R^{-1}\sqrt{Q_F} = UD^2U^\dag \\
    e^{-\sigma[Q_R,Q_F]} &= Q_R^{-1/2}Q_F Q_R^{-1/2} = VD^2V^\dag 
\end{align}
and therefore, $\sigma[Q_F,Q_R^\gamma] = 2U(\log D) U^\dag$ and $\sigma[Q_R,Q_F] =-2V(\log D) V^\dag$. This also verifies that $\sigma[Q_F,Q_R]$ and $\sigma[Q_R,Q_F]$ have opposite eigenvalues and $s_k = 2\log d_k$.

From \cref{eq:singular-decomp} and that $Q_F,Q_R$ are Hermitian, one gets $Q_F = UDV^\dag Q_R VDU^\dag$.
From \cref{eq:UVS}, one gets $\ket{f_k} = UV^\dag\ket{r_k}$. From these, for any $k$,
\begin{align}
    \mathbb{P}_F(s_k) &= \bra{f_k}Q_F\ket{f_k}\\
    &= \bra{r_k}VU^\dag \times UDV^\dag Q_R VDU^\dag \times UV^\dag \ket{r_k} \\
    & = \bra{r_k}VDV^\dag Q_R VDV^\dag \ket{r_k} \\
    & =  d_k^2 \bra{r_k} Q_R \ket{r_k}  \\
    & =  d_k^2~ \mathbb{P}_R(-s_k) \label{eq:QFskQR}
\end{align}
noting that $\ket{r_k}$ is the eigenvector of $VDV^\dag$ and $VDV^\dag\ket{r_k} = d_k\ket{r_k}$. Last, since $s_k = 2\log d_k$, one gets $\mathbb{P}_R(-s_k) = e^{-s_k} \mathbb{P}_F(s_k)$, which is the detailed fluctuation theorem in \cref{crooksc}.

\section{Physical interpretation of $Q_F$ and $Q_R^\gamma$}
\label{app:interpretation_tQ}

By Stinespring's dilation theorem \cite{stinespring1955positive}, any channel can be written as an isometric channel from $A$ to a composite system $BE$ consisting of the output $B$ and an environment system $E$, followed by tracing out system $E$:
\begin{align}
    \mathcal{E}(\rho) = \Tr_E[V \rho V^\dag]
\end{align}
where $V$ is an isometry from $\HS_A$ to $\HS_B\otimes\HS_E$. The complementary channel \cite{devetak2005capacity} is obtained by tracing out the original output system $B$, while keeping the environment system $E$:
\begin{align}
    \mathcal{E}^c(\rho) \coloneqq \Tr_B[V\rho V^\dag]\,.
\end{align}
The complementary channel depends on the isometry $V$ used in the dilation, but is unique up to an isometry on the environment system \cite{king2005properties,holevo2007complementary}.

We first show that $Q_F$ is a complementary channel of $\map{E}$.
Choose $\HS_E = \HS_{B'}\otimes\HS_{A'}$ with $\HS_{A'}\cong\HS_A$ and $\HS_{B'}\cong\HS_B$. Let $\one_{A'\gets A}$ be the isomorphism from $A$ to $A'$ and $\one_{A\gets A'}$ be its inverse, and $\one_{B'\gets B}$ be the isomorphism from $B$ to $B'$ and $\one_{B\gets B'}$ be its inverse. 
In the following, we use $\rho_{A'}\coloneqq \one_{A'\gets A}\rho\one_{A\gets A'}$ to denote the operator on system $A'$ corresponding to $\rho$. Similar notations are also used for other operators such as $\mathcal{E}(\gamma)_{B'}\coloneqq\one_{B'\gets B} \mathcal{E}(\gamma)\one_{B\gets B'}$ and $\tau_{B'} \coloneqq \one_{B'\gets B} \tau \one_{B\gets B'}$. Define
\begin{align}
    V\coloneqq \left(\one_B\otimes\sqrt{C_\mathcal{E}^T}\right)(\ket{\Phi^+}_{BB'}\otimes \one_{A'\gets A}) : \HS_A \to \HS_B\otimes\HS_E
\end{align}
where $C_\mathcal{E}$ is considered as an operator on system $E$ (namely $B'A'$),  and $\ket{\Phi^+}_{B'B}\coloneqq\sum_{i}\ket{i}_{B'}\ket{i}_B$ is the unnormalized maximally entangled state on systems $B'B$. This is an isometry because
\begin{align}
    V^\dag V = (\bra{\Phi^+}_{BB'}\otimes \one_{A\gets A'})(\one_B\otimes{C_\mathcal{E}^T})(\ket{\Phi^+}_{BB'}\otimes \one_{A'\gets A}) = \one_{A\gets A'}\Tr_{B'}[C_\mathcal{E}]\one_{A'\gets A} = \one_A \,.
\end{align}

We will see that with this definition, up to isomorphisms between $A$ and $A'$ and between $B$ and $B'$,
\begin{align}
    \mathcal{E}(\rho) &= \Tr_E[V \rho V^\dag] \label{eq:VrhoVdag=E_rho} \,,\\
    Q_F^T= Q_F(\rho)^T = \mathcal{E}^c(\rho) &\coloneqq \Tr_B[V\rho V^\dag]\,. \label{eq:VrhoVdag=Q_F}
\end{align}
\cref{eq:VrhoVdag=E_rho} implies that $V$ is a correct dilation of $\mathcal{E}$, and \cref{eq:VrhoVdag=Q_F} shows that $Q_F$ can be viewed as a complementary channel of $\map{E}$ up to a transpose. 
\begin{align}
    \Tr_E[V\rho V^\dag] &= \Tr_E\left[(\one_B \otimes \sqrt{C_\mathcal{E}^T})(\ketbrasame{\Phi^+}_{BB'}\otimes\rho_{A'}) (\one_B \otimes \sqrt{C_\mathcal{E}^T})\right]\\
    & = \Tr_{E}\left[(\one_B \otimes {C_\mathcal{E}^T})(\ketbrasame{\Phi^+}_{BB'}\otimes\rho_{A'})\right]\\
    & = \Tr_{B'}\left[\ketbrasame{\Phi^+}_{BB'}\left(\one_B \otimes \Tr_{A'}\left[{C_\mathcal{E}^T}(\one_{B'} \otimes\rho_{A'})\right]\right)\right]\\
    &=\one_{B\gets B'}\Tr_{A'}\left[{C_\mathcal{E}^T}(\one_{B'} \otimes\rho_{A'})\right]^T \one_{B'\gets B} \label{eq:use_Phi+_B}\\
    & = \mathcal{E}(\rho) 
\end{align}
In \cref{eq:use_Phi+_B}, we have used $\Tr_B[\ketbrasame{\Phi^+}_{BB'}(\one_B\otimes X)] = \one_{B\gets B'} X^T \one_{B'\gets B}$, where the transpose is defined on the basis $\{\ket{i}_{B'}\}$.  Now we compute the marginal state on the environment system.
\begin{align}
    \Tr_B[V\rho V^\dag] &= \Tr_B\left[(\one_B \otimes \sqrt{C_\mathcal{E}^T})(\ketbrasame{\Phi^+}_{BB'}\otimes\rho_{A'}) (\one_B \otimes \sqrt{C_\mathcal{E}^T})\right]\\
    & = \sqrt{C_\mathcal{E}^T}(\one_{B'}\otimes\rho_{A'})\sqrt{C_\mathcal{E}^T} \\
    & = Q_F(\rho)^T_{B'A'}
\end{align}
This shows \cref{eq:VrhoVdag=Q_F}.

Next, we show that $Q_R$ is a complementary map of the reverse process $\mathcal{R}_{\mathcal{E}}^\gamma$. We define the following isometry
\begin{align}
    V\coloneqq (\sqrt{C_\mathcal{E}}\otimes \sqrt{\gamma} )(\one_{B'\gets B}\mathcal{E}(\gamma)^{-1/2} \otimes \ket{\Phi^+}_{A'A}) : \HS_B \to \HS_E\otimes\HS_A
\end{align}
and will show that
\begin{align}
    \mathcal{R}_{\mathcal{E}}^\gamma(\tau) &= \Tr_E[V \tau V^\dag] \label{eq:VtauVdag=R_tau} \,,\\
    Q_R^\gamma(\tau) =(\mathcal{R}_{\mathcal{E}}^\gamma)^c(\tau) &\coloneqq \Tr_A[V\tau V^\dag]\,. \label{eq:VtauVdag=Q_R}
\end{align}

First, $V$ is an isometry because
\begin{align}
    V^\dag V &= (\mathcal{E}(\gamma)^{-1/2}\one_{B\gets B'} \otimes \bra{\Phi^+}_{A'A})({C_\mathcal{E}}\otimes{\gamma} )(\one_{B'\gets B}\mathcal{E}(\gamma)^{-1/2} \otimes \ket{\Phi^+}_{A'A}) \\
    & = \mathcal{E}(\gamma)^{-1/2}\one_{B\gets B'} \Tr_{A'}\left[{C_\mathcal{E}}(\one_{B'}\otimes{\gamma}^T)\right]\one_{B'\gets B}\mathcal{E}(\gamma)^{-1/2} \\
    & = \mathcal{E}(\gamma)^{-1/2}\mathcal{E}(\gamma)\mathcal{E}(\gamma)^{-1/2} \\
    & = \one_B
\end{align}

Second,
\begin{align}
    \Tr_E[V \tau V^\dag] &= \Tr_E\left[(\sqrt{C_\mathcal{E}}\otimes \sqrt{\gamma} )\left(\mathcal{E}(\gamma)^{-1/2}_{B'} \tau_{B'} \mathcal{E}(\gamma)^{-1/2}_{B'} \otimes \ketbrasame{\Phi^+}_{A'A}\right)(\sqrt{C_\mathcal{E}}\otimes \sqrt{\gamma} ) \right] \\
    & = \sqrt{\gamma} \, \Tr_E\left[({C_\mathcal{E}}\otimes \one_A )\left(\mathcal{E}(\gamma)^{-1/2}_{B'} \tau_{B'} \mathcal{E}(\gamma)^{-1/2}_{B'} \otimes \ketbrasame{\Phi^+}_{A'A}\right)\right]\sqrt{\gamma} \\
    & = \sqrt{\gamma} \, \Tr_{A'}\left[ \left(\Tr_{B'}\left[{C_\mathcal{E}}\left(\mathcal{E}(\gamma)^{-1/2}_{B'} \tau_{B'} \mathcal{E}(\gamma)^{-1/2}_{B'}\otimes\one_{A'} \right)\right]\otimes \one_A \right)  \otimes \ketbrasame{\Phi^+}_{A'A}\right]\sqrt{\gamma} \\
    & = \sqrt{\gamma} \, \Tr_{A'}\left[ \left(\map{E}^\dag\left[\mathcal{E}(\gamma)^{-1/2} \tau\mathcal{E}(\gamma)^{-1/2}\right]^T_{A'} \otimes \one_A \right)  \otimes \ketbrasame{\Phi^+}_{A'A}\right]\sqrt{\gamma} \label{eq:channel_dagger_use} \\
    & = \sqrt{\gamma} \, \map{E}^\dag\left[\mathcal{E}(\gamma)^{-1/2} \tau \mathcal{E}(\gamma)^{-1/2}\right]\sqrt{\gamma}\\
    & = \map{R}_{\map{E}}^\gamma(\tau)
\end{align}
which shows \cref{eq:VtauVdag=R_tau}. In \cref{eq:channel_dagger_use}, we have used $\Tr_{B}[C_{\mathcal{E}}(\sigma\otimes\one_A)] = \mathcal{E}^\dag(\sigma)^T$.
Last,
\begin{align}
    \Tr_A[V \tau V^\dag] &= \Tr_A\left[(\sqrt{C_\mathcal{E}}\otimes \sqrt{\gamma} )\left(\mathcal{E}(\gamma)^{-1/2}_{B'} \tau_{B'} \mathcal{E}(\gamma)^{-1/2}_{B'} \otimes \ketbrasame{\Phi^+}_{A'A}\right)(\sqrt{C_\mathcal{E}}\otimes \sqrt{\gamma} ) \right] \\
    & = \sqrt{C_\mathcal{E}} (\mathcal{E}(\gamma)^{-1/2}_{B'} \tau_{B'} \mathcal{E}(\gamma)^{-1/2}_{B'}\otimes\one_{A'}) \,\Tr_A\left[(\one_{A'}\otimes {\gamma} ) \ketbrasame{\Phi^+}_{A'A} \right] \sqrt{C_\mathcal{E}} \\
    & = \sqrt{C_\mathcal{E}} (\mathcal{E}(\gamma)^{-1/2}_{B'} \tau_{B'} \mathcal{E}(\gamma)^{-1/2}_{B'}\otimes\one_{A'}) \gamma^T_{A'} \sqrt{C_\mathcal{E}} \\
    & = \sqrt{C_\mathcal{E}} (\mathcal{E}(\gamma)^{-1/2}_{B'} \tau_{B'} \mathcal{E}(\gamma)^{-1/2}_{B'}\otimes\gamma^T_{A'})  \sqrt{C_\mathcal{E}} \\
    & = Q_R^\gamma(\tau)_{B'A'}
\end{align}
This shows \cref{eq:VtauVdag=Q_R}.

\section{Proof of \cref{avggen}}
\label{app:avggen}

We start with \cref{locality} and use the property that $U Q_F U^\dag=\tilde{Q}_F$, with $\tilde{Q}_F$ defined in \cref{eq:QFtilde}. For all well-behaved (analytical) functions $f$ such as e.g., the moment generating function, we have $P^{-1}f(X)P = f(P^{-1}XP)$ for invertible operator $P$. Using this and cyclic property of the trace, we have
\begin{align}
    &\textrm{Tr}\left[Q_F\ f(\sigma[Q_F,Q_R^\gamma])\right] \nonumber\\
    &= \Tr\left[\sqrt{Q_F} Q_F Q_F^{-1/2}\   f\left(   \log \left\{\sqrt{Q_F}\,Q_R^{-1}\,\sqrt{Q_F}\right\} \right) \right]\\
    & = \Tr\left[Q_F\ f\left(   \log \left\{Q_F^{-1/2} \sqrt{Q_F}\,Q_R^{-1}\,\sqrt{Q_F} \sqrt{Q_F} \right\} \right) \right]\\
    & = \Tr\left[Q_F\ f\left(   \log Q_R^{-1} Q_F  \right) \right]\\
    & = \Tr\left[Q_F\ f\left(   \log \left[ C_{\mathcal{E}}^{-1/2}\left(\mathcal{E}(\gamma)^{-1/2}\tau \mathcal{E}(\gamma)^{-1/2}\otimes \gamma^T\right)^{-1}(\one\otimes\rho^T)\sqrt{C_{\mathcal{E}}} \right]\right) \right]\\
    & = \Tr\left[\sqrt{C_{\mathcal{E}}} Q_FC_{\mathcal{E}}^{-1/2}\ f\left(   \log \left[ \left(\mathcal{E}(\gamma)^{-1/2}\tau \mathcal{E}(\gamma)^{-1/2}\right)^{-1}\otimes (\gamma^T)^{-1} \rho^T\right]\right) \right]\\
    & = \Tr\left[\sqrt{C_{\mathcal{E}}} Q_FC_{\mathcal{E}}^{-1/2}\ f\left(   \log \left[ \left(\mathcal{E}(\gamma)^{-1/2}\tau \mathcal{E}(\gamma)^{-1/2}\right)^{-1}\otimes (\sqrt{\rho^T})^{-1} \sqrt{\rho^T}(\gamma^T)^{-1} \sqrt{\rho^T}\sqrt{\rho^T}\right]\right) \right]\\
    & = \Tr\left[(\one\otimes\sqrt{\rho^T})\sqrt{C_{\mathcal{E}}} Q_FC_{\mathcal{E}}^{-1/2}(\one\otimes\sqrt{\rho^T})^{-1} \ f\left(   \log \left[ \left(\mathcal{E}(\gamma)^{-1/2}\tau \mathcal{E}(\gamma)^{-1/2}\right)^{-1}\otimes \sqrt{\rho^T}(\gamma^T)^{-1} \sqrt{\rho^T}\right]\right) \right]\\
    & = \Tr\Big[\tilde{Q}_F\ f\big(\one\otimes \log \sqrt{\rho^T}(\gamma^T)^{-1}\sqrt{\rho^T} - \log \mathcal{E}(\gamma)^{-1/2}\tau\mathcal{E}(\gamma)^{-1/2} \otimes \one \big)\Big]\;.\label{eq:C_E_cancel}
\end{align}

Taking $f$ to be the identity function $f(x)=x$, using $\textrm{Tr}_A[\tilde{Q}_F(\rho)]=\rho^T$ and $\textrm{Tr}_B[\tilde{Q}_F(\rho)]=\mathcal{E}(\rho)$, we obtain
\begin{align}
    \Sigma &=\textrm{Tr}\left[Q_F\ \sigma[Q_F,Q_R^\gamma]\right]\\
    &=\Tr\Big[\tilde{Q}_F\ \big(\one\otimes \log \sqrt{\rho^T}(\gamma^T)^{-1}\sqrt{\rho^T} - \log \mathcal{E}(\gamma)^{-1/2}\tau\mathcal{E}(\gamma)^{-1/2} \otimes \one \big)\Big]\\
    & = \Tr\Big[\tilde{Q}_F\ \big(\one\otimes \log \sqrt{\rho^T}(\gamma^T)^{-1}\sqrt{\rho^T}\big)\Big] - \Tr\Big[\tilde{Q}_F \big( \log \mathcal{E}(\gamma)^{-1/2}\tau\mathcal{E}(\gamma)^{-1/2} \otimes \one \big)\Big] \\
    & = \Tr\Big[\rho^T \log \sqrt{\rho^T}(\gamma^T)^{-1}\sqrt{\rho^T}\Big] - \Tr\Big[\mathcal{E}(\rho) \log \mathcal{E}(\gamma)^{-1/2}\tau\mathcal{E}(\gamma)^{-1/2} \Big] \\
    &=D_{\textrm{BS}}(\rho\|\gamma)-\textrm{Tr}\left[\mathcal{E}(\rho)\log\left(\mathcal{E}(\gamma)^{-1/2}\tau \mathcal{E}(\gamma)^{-1/2}\right)\right]\,.
\end{align}

\section{Proof of \cref{eq:tau=Erho}}
\label{app:tau=Erho}

The main tool we use to prove inequalities is the following result in Ref.~\cite{carlen2019some}:
\begin{lemma} \label{lem:tracelogineq}
    For positive operators $X,Y,Z$, with $\Tr[X]=\Tr[Z]$, and all $p>0$,
    \begin{align}
        \Tr[X\log(Y^{p/2}Z^pY^{p/2})] \leq \Tr[X(\log X^p +\log Y^p)] \leq \Tr[X\log(X^{p/2}Y^pX^{p/2})]
    \end{align}
    The first inequality is strict unless $[Y,Z]=0$, and the second inequality becomes equality if and only if $[X,Y]=0$.
\end{lemma}

We use this to prove \cref{eq:tau=Erho}. By \cref{lem:tracelogineq} with $X=Z=\mathcal{E}(\rho)$, $Y=\mathcal{E}(\gamma)^{-1}$  and $p=1$,
\begin{align}
    \textrm{Tr}\left[\mathcal{E}(\rho)\log\left(\mathcal{E}(\gamma)^{-1/2}\mathcal{E}(\rho)\mathcal{E}(\gamma)^{-1/2}\right)\right] \leq \Tr\left[\mathcal{E}(\rho) (\log\mathcal{E}(\rho) - \log\mathcal{E}(\gamma) \right] = D(\mathcal{E}(\rho)\|\mathcal{E}(\gamma)) \,.
\end{align}
With this inequality, for $\tau=\mathcal{E}(\rho)$,
\begin{align}
\avg{\sigma[Q_F,Q_R^\gamma]}_F&=D_{\textrm{BS}}(\rho\|\gamma)-\textrm{Tr}\left[\mathcal{E}(\rho)\log\left(\mathcal{E}(\gamma)^{-1/2}\mathcal{E}(\rho)\mathcal{E}(\gamma)^{-1/2}\right)\right]\\
&\geq D_{\textrm{BS}}(\rho\|\gamma)- D(\mathcal{E}(\rho)\|\mathcal{E}(\gamma))
\end{align}

\section{Superadditivity of average entropy production}
\label{app:superadditivity}

By \cref{avggen},
\begin{align}
&\Sigma_1+\Sigma_2 -\Sigma_{12} \\
&= D_{\textrm{BS}}(\rho\|\gamma)-\textrm{Tr}\left[\mathcal{E}_1(\rho)\log\left(\mathcal{E}_1(\gamma)^{-1/2}\tau_1 \mathcal{E}_1(\gamma)^{-1/2}\right)\right]\nonumber \\
&\mathrel{\phantom=} + D_{\textrm{BS}}(\mathcal{E}_1(\rho)\|\mathcal{E}_1(\gamma))-\textrm{Tr}\left[\mathcal{E}_2[\mathcal{E}_1(\rho)]\log\left(\mathcal{E}_2[\mathcal{E}_1(\gamma)]^{-1/2}\tau_2 \mathcal{E}_2[\mathcal{E}_1(\gamma)]^{-1/2}\right)\right] \nonumber \\
&\mathrel{\phantom=} -\left( D_{\textrm{BS}}(\rho\|\gamma)-\textrm{Tr}\left[\mathcal{E}_2[\mathcal{E}_1(\rho)]\log\left(\mathcal{E}_2[\mathcal{E}_1(\gamma)]^{-1/2}\tau_2 \mathcal{E}_2[\mathcal{E}_1(\gamma)]^{-1/2}\right)\right] \right) \\
& = D_{\textrm{BS}}(\mathcal{E}_1(\rho)\|\mathcal{E}_1(\gamma))- \textrm{Tr}\left[\mathcal{E}_1(\rho)\log\left(\mathcal{E}_1(\gamma)^{-1/2}\tau_1 \mathcal{E}_1(\gamma)^{-1/2}\right)\right] \label{eq:1+2-12}
\end{align}

The last line is no less than zero by taking $X=\mathcal{E}_1(\rho)$, $Y=\mathcal{E}_1(\gamma)^{-1}$, $Z=\tau_1$ and $p=1$ in \cref{lem:tracelogineq}.

This lemma also indicates that, for \cref{eq:1+2-12} to be zero, one needs $[\mathcal{E}_1(\rho),\mathcal{E}_1(\gamma)]=[\mathcal{E}_1(\gamma),\tau_1]=0$, in which condition \cref{eq:1+2-12} becomes $D(\mathcal{E}_1(\rho)\|\tau_1)$, which is zero if and only if $\tau_1=\mathcal{E}_1(\rho)$. Therefore, $\Sigma_1+\Sigma_2 -\Sigma_{12}$ if and only if $\tau_1=\mathcal{E}_1(\rho)$ and $[\mathcal{E}_1(\rho),\mathcal{E}_1(\gamma)]=0$.

\end{document}